\documentclass[usenatbib,usegraphicx]{mn2e}
\usepackage{amssymb}

\title[A survey of the $\eta$ Chamaeleontis cluster to the brown
       dwarf -- planet boundary]
      {A deep photometric survey of the $\eta$ Chamaeleontis cluster
       down to the brown dwarf -- planet boundary}

\author[A-R. Lyo et al.]
{A-Ran Lyo,$^{1}$\thanks{E-mail:
arl@asiaa.sinica.edu.tw (ARL);
song@gemini.edu (IS);
w.lawson@adfa.edu.au (WAL);
bessell@mso.anu.edu.au (MB);
ben@astro.ucla.edu (BZ)}
Inseok Song,$^{2}$ Warrick A. Lawson,$^{3}$ M. S. Bessell$^{4}$
and B. Zuckerman$^{5}$\\
$^{1}$Academia Sinica Institute of Astronomy and Astrophysics,
PO Box 23-141, Taipei 106, Taiwan\\
$^{2}$Gemini Observatory, 670 North A'ohoku Place, Hilo, HI 96720, USA\\
$^{3}$School of Physical, Environmental and Mathematical Sciences,
University of New South Wales, Australian Defence Force Academy,\\
Canberra, ACT 2600, Australia\\
$^{4}$Research School of Astronomy and Astrophysics, Institute of
Advanced Studies, The Australian National University, Cotter Road,\\
Weston Creek ACT 2611, Australia\\
$^{5}$Department of Physics and Astronomy and Center for Astrobiology,
University of California, Los Angeles, CA 90095-1562, USA}

\begin{document}

\date{Accepted .................... Received ....................}

\pagerange{\pageref{firstpage}--\pageref{lastpage}}
\pubyear{2005}
\maketitle

\label{firstpage}

\begin{abstract}
We report the outcome of the deep optical/infrared photometric
survey of the central region ($33 \times 33$ arcmin or 0.9
pc$^{2}$) of the $\eta$ Chamaeleontis pre-main sequence star
cluster.  The completeness limits of the photometry are $I =
19.1$, $J = 18.2$ and $H = 17.6$; faint enough to reveal low mass
members down to the brown dwarf and planet boundary of $\approx
13$ M$_{Jup}$.  We found no such low mass members in this region.
Our result combined with a previous shallower ($I = 17$) but
larger area survey indicates that low mass objects ($0.013 <
M$/M$_{\odot} < 0.075$) either were not created in the $\eta$ Cha
cluster or were lost due to the early dynamical history of the
cluster and ejected to outside the surveyed areas.
\end{abstract}


\section{Introduction}

\begin{figure*}
\begin{center}
\includegraphics[width=170mm]{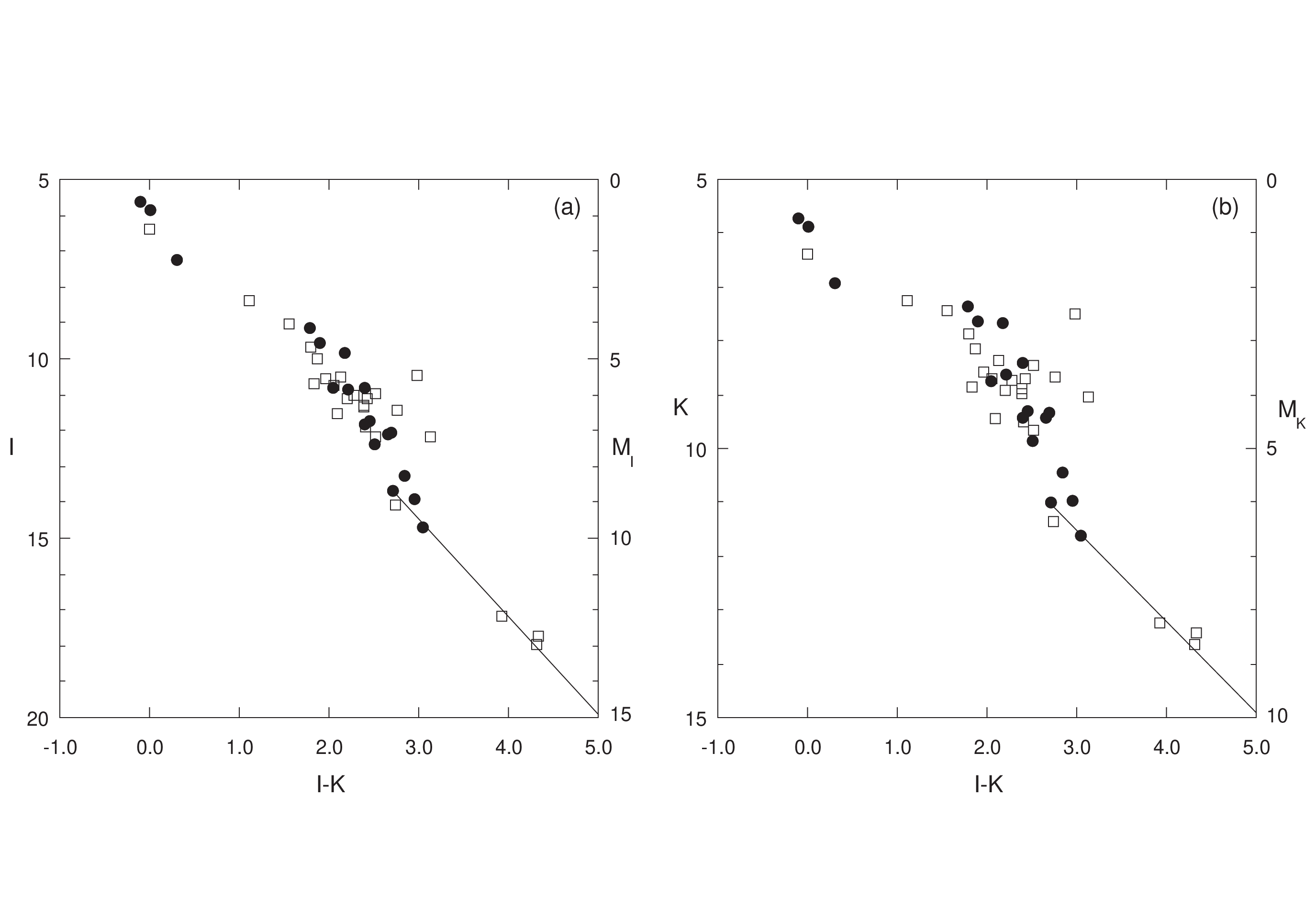}
%\vspace{1cm}
\caption{(a) $I$ versus $(I-K)$ and (b) $K$ versus
$(I-K)$ colour-magnitude diagrams of the known members of the
$\eta$ Cha cluster and TW Hydrae association. The right-hand-side
axes are marked in absolute magnitudes. The filled-circles are for
the members of the $\eta$ Cha cluster, while the open-squares are
for members of the TWA if placed at the $\eta$ Cha distance of $d
= 97$ pc (see Mamajek 2005 for distances to the TWA members). The
scatter in magnitude at any given colour is principally a function
of unaccounted for multiplicity. The straight solid line in each
graph represents a $\sim 10$ Myr (the approximate age of the
$\eta$ Cha and TWA groups) pseudo-isochrone for very low-mass
stars and brown dwarfs to obtain the relationship between $I$- and
$K$-band luminosity; see Section 1.}
%\label{Fig1}
\end{center}
\end{figure*}

The nature of the initial mass function (IMF) for brown dwarfs and
their spatial distribution hold important keys to the dominant brown
dwarf formation mechanism in star clusters.  This will provide a
solution to the formation mechanism of brown dwarfs; whether brown
dwarfs are stellar embryos, ejected from unstable newborn multiple
systems before they can accrete sufficient mass for hydrogen fusion
(Papaloizou \& Terquem 2001; Reipurth \& Clarke 2001; Bate et al.
2003; Delgado-Donate et al. 2003; Kroupa \& Bouvier 2003; Sterzik
\& Durisen 2003); or hydrostatic cores that lose their accretion
envelopes because of encounters with other protostars in rich
clusters (Price \& Podsiadlowski 1995); or by photo-evaporation of
their accretion envelopes through nearby O stars (Kroupa et al. 1999;
Matsuyama et al. 2003; Whitworth \& Zinnecker 2004); or if they form
in the same way as stars (Shu et al. 1987; Briceno et al. 2002;
White \& Basri 2003; Padoan \& Nordlund 2004).

The recently discovered $\eta$ Chamaeleontis cluster ($\eta$ Cha
hereafter; Mamajek et al. 1999) is an ideal target to study the
lowest mass objects due to its youth ($t \simeq 8$ Myr, Lawson \&
Feigelson 2001; Zuckerman \& Song 2004), compact size (extent
$\sim 1$ pc), and proximity to Earth ($d \sim 97$ pc). Negligible
foreground reddening has been determined for $\eta$ Cha [HD 75416,
$E(b-y) = -0.004$; Westin 1985] simplifying photometric analysis.
These characteristics of the $\eta$ Cha cluster enable
conventional photometric surveys with medium size telescopes
($3-4$ m aperture) to detect candidate members even in the
planetary mass range.  In addition, the compactness of the cluster
compared to other nearby young stellar associations is an
advantage for attempting to determine complete cluster membership
down to very low masses. On the contrary, a deep photometric
membership survey of the TW~Hydrae association (TWA) (size $> 500
{\rm deg}^2$ in the sky plane), which like the $\eta$ Cha cluster
also appears to share a common kinematic link with the subgroups
of the Ophiuchus-Scorpius-centaurus OB association (Mamajek et al.
2000), is not practical and likely contains too many field
interlopers.

To date, 18 primaries have been discovered in the $\eta$ Cha
cluster with a mass range from 0.15 M$_{\odot}$ to 3.4 M$_{\odot}$
(Mamajek et al. 1999; Lawson et al. 2002; Lyo et al. 2004; Song et
al. 2004; Luhman \& Steeghs 2004).  By extending the cluster IMF
to lower masses, Lyo et al. (2004) predicted that $10-14$
additional low mass stars with $M$/M$_{\odot} = 0.08-0.15$ and
$10-15$ brown dwarfs with $M$/M$_{\odot} = 0.025-0.08$ remain to
be discovered, a number of objects comparable to the already known
stellar population. An alternative calculation is to use the IMF
relationships for the mass of the stellar membership of a cluster
put forward by Weidner \& Kroupa (2006; see their figs 1 and 6).
For the $\eta$ Cha cluster, the Weidner \& Kroupa relations also
predict a total population greatly in excess of the known
population, with $\approx 90$ objects with $M > 0.01$M$_{\odot}$.

In a recent search for low mass members of the $\eta$ Cha cluster,
Luhman (2004) found no new members within a radius of
1.5$^{\circ}$ surrounding $\eta$ Cha using DEep Near Infrared
Survey (DENIS) and Two Micron All Sky Survey (2MASS) photometric
data.  Using Baraffe et al. (1998) and Chabrier et al. (2000)
pre-main sequence tracks and a modified
luminosity-colour-temperature sequence, Luhman claimed his survey
was complete (at the $i = 17$ completeness limit of the DENIS
survey) for low mass objects within a mass range of $M$/M$_{\odot}
= 0.015-0.15$.  However, the actual mass range is very sensitive
to the adopted isochrone, which has been poorly defined by
observations in the brown dwarf regime for $\sim 10$ Myr-old
objects.  Recently, Mamajek (2005) has calculated kinematic
distances to members of the TWA which has a similar age ($\sim 10$
Myr) to that of the $\eta$ Cha cluster. The TWA includes three
brown dwarfs with masses of $20-25$ M$_{Jup}$ determined from the
DUSTY tracks of Chabrier et al. (2000) and other lines of
evidence.  These three objects are the $\sim 25$ M$_{Jup}$ 2MASSW
J1207334-393254 and 2MASSW J1139511-315921 (Gizis 2002) and the
$\sim 20$ M$_{Jup}$ SSSPM J1102-3431 (Scholz et al. 2005). If
these objects are placed at the distance of the $\eta$ Cha
cluster, they all fall below the $i = 17$ mag limit of the DENIS
survey, suggesting that a DENIS-based survey of the $\eta$ Cha
cluster is limited to $M \gtrsim 25$ M$_{Jup}$ objects.

We also obtain this same mass limit directly from the
mass-luminosity relation from the DUSTY models (Chabrier et al.
2000), making use of the absolute $K$-band magnitudes for the TWA
brown dwarfs.  Using the $K$-band luminosity to obtain the mass is
acceptable as most of the flux in young very low-mass objects is
released in the near-infrared (the peak flux of the lowest-mass
objects arises at $\sim 1 \mu$m). Recently, Close et al. (2005)
reported that the predicted mass of AB Dor C from the $K$-band
flux (0.070 M$_{\odot}$) compared reasonably well to their
dynamical mass of $0.090 \pm 0.005$ M$_{\odot}$ obtained through
direct imaging, even if it is still a significant (20 per cent)
underestimate.  (Close et al. also warn about the danger of using
evolutionary tracks at ages that have not been sufficiently
calibrated by observations).  Fig. 1(a) shows the $I$ versus
$(I-K)$ colour-magnitude diagram for the known members of the
$\eta$ Cha cluster and the TWA. We plotted the members of the TWA
as if they were located at the $d = 97$ pc distance to the $\eta$
Cha cluster (see Mamajek 2005 for kinematic distances to the TWA
members).  Together these observations map the $\sim 10$ Myr
isochrone across spectral types ranging from B8 ($\eta$ Cha
itself; $M = 3.4$ M$_{\odot}$) to the three M8 -- M8.5 TWA brown
dwarfs with masses of $20-25$ M$_{\it Jup}$. The straight solid
lines in Figs 1(a) and 1(b) approximately map the isochrone across
the brown dwarf mass regime in these colour-magnitude diagrams.
From this observational isochrone, we conclude the mass limit of
the DENIS survey to be $\sim 25$ M$_{Jup}$ using the observed mass
and K-band magnitude relation (Fig. 1b), in conjunction with the 5
Myr and 10 Myr DUSTY models (also see Section 2 for further
discussion of this calibration).

In this paper, we report the result of the deepest
optical/infrared photometric survey attempted on the $\eta$ Cha
cluster to date.  The new observations that we report here extend
the DENIS and 2MASS surveys by $\simeq 2$ magnitudes.  We show
that our survey is sensitive to low-mass objects near the brown
dwarf -- planet boundary with $M \approx 0.013$ M$_{\odot}$.

\section{Observations and data reduction}

\begin{figure*}
\begin{center}
\includegraphics[width=170mm]{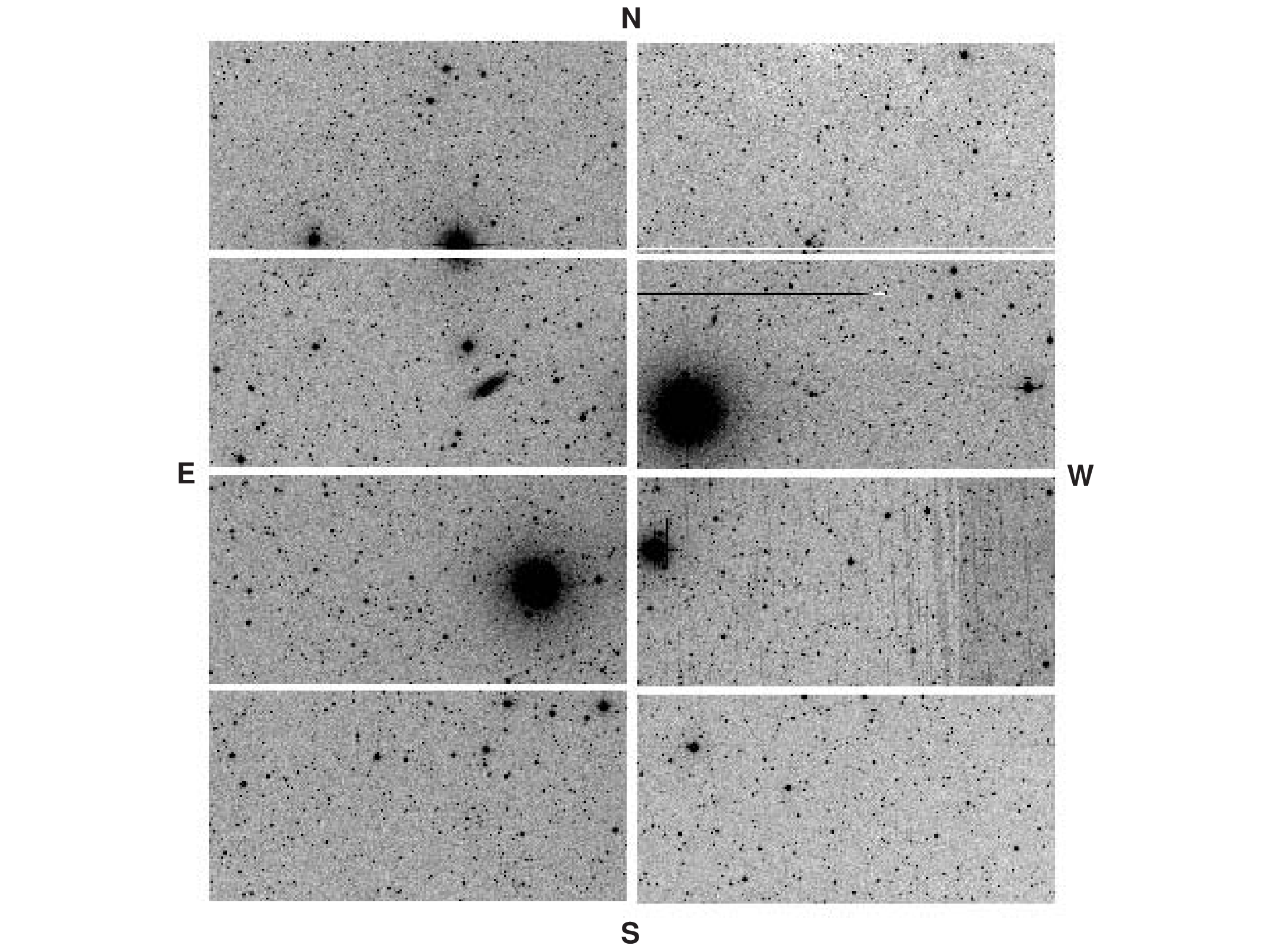}
%\vspace{-3cm}
\caption{$33 \times 33$ arcmin WFI $I$-band image of
the $\eta$ Cha cluster core region centered at $\alpha$, $\delta$
(2000) = 8$^{\rm h}$41$^{\rm m}$59.9$^{\rm s}$,
$-79^{\circ}$00$^{'}$00$^{"}$; the spatial center of the three
early-type systems in the cluster. The combined coverage of the
IRIS2 $JH$-band images is almost the same as the $I$-band field;
see Section 2 for details.}
%\label{Fig2}
\end{center}
\end{figure*}

An $I$-band image of the $\eta$ Cha cluster was obtained in
service mode using the 3.9-m Anglo-Australian Telescope (AAT) and
Wide Field Imager (WFI) on 2004 January 20. WFI is an 8k $\times$
8k CCD mosaic (eight 2048 $\times$ 4096 15-$\mu$m square pixel
MIT/LL CCDs) for optical imaging at the prime focus. The image
scale is 0.23 arcsec pixel$^{-1}$ at f/3.3, covering a $33 \times
33$ arcmin field (Fig. 2). The exposure time was 60 sec and the
seeing was $\sim 1.5$ arcsec.  The WFI $I$-band image was overscan
corrected, bias and dark subtracted, linearized, and flat-fielded
using the {\tt IRAF/MSCRED} package. We then split this mosaic
frame into eight individual CCD images using {\tt MSCRED/mscsplit}
for further photometric reduction.

$JH$-band images of almost the same field as the WFI $I$-band image
were obtained on 2004 April 5 in service mode with the AAT using
the $1.0-2.5 \mu$m infrared imager and longslit/multi-slit spectrograph
(IRIS2).  IRIS2 uses a $1024 \times 1024$ Rockwell HAWAII-1 HgCdTe
infrared detector which has a plate scale of 1.45 arcsec pixel$^{-1}$,
giving a field of $7.7 \times 7.7$ arcmin at the f/8 Cassegrain focus.
We obtained 22 images with $3 \times 9$-s exposure cycles (27-s total
exposure time at each position) in each filter covering the central
$\sim 30 \times 30$ arcmin region of the cluster.  Unfortunately,
the seeing was quite variable in each of the image fields, ranging
between 1.5 and 3.0 arcsec.  However, conditions remained photometric.
The raw images were pre-processed using the ORAC-DR data reduction
pipeline (Economou et al. 1998).

Instrumental magnitudes from the WFI $I$-band and IRIS2 $JH$-band
data were obtained using the {\tt IRAF/DAOPHOT\,} task via
point-spread function fitting.  For the calibration of the
$I$-band photometry, we used photometry from the DENIS {\it
Gunn-i\,} (0.82 $\mu$m) survey. For the $JH$-band data, we used
2MASS $J$ (1.25 $\mu$m) and $H$ (1.65 $\mu$m) photometry.  We
selected well-isolated objects with DENIS and 2MASS data to
transform instrumental into real magnitudes using independent
calibrations for each frame and filter.  We did not consider
colour terms in these transformations because we have only one
optical band and the $JH$-band data did not show any definite
colour term.  For the WFI $I$-band observations, the scatter in
these magnitude transformations (to check the accuracy of the
magnitudes) was 0.07 mag (0.16 mag) for stars brighter (fainter)
than $I = 17$. For the IRIS2 $J-$ and $H-$band observations, the
scatter was typically 0.06 mag (0.08 mag) for stars brighter
(fainter) than $\sim 16$ mag at $J$ or $H-$band. Scatters are
slightly larger for fields with the poorest seeing. However, such
scatter has little influence on our results as we show in Fig. 3.

Fig. 3 shows the WFI/IRIS2 $(I-J)$ versus $I$ colour-magnitude
diagram, with our survey photometry shown as plus symbols. To show
the diagram space occupied by cluster members and any candidate
members, we overplot the known primaries.  The filled-circles are
photometry for the 18 known primaries of the $\eta$ Cha cluster,
while the open-squares are photometry for the members of the TWA
after correcting their magnitudes to the $\eta$ Cha distance
(Mamajek 2005).  Late-type stars and brown dwarfs associated with
the $\eta$ Cha cluster will be clearly elevated in magnitude above
the vast majority of field stars.  The typical uncertainties in
our survey photometry can be neglected compared to the wide range
in magnitude and colour plotted in Fig. 3.

The two long-dashed lines in Fig. 3 are the 5 Myr ($0.008-0.075$
M$_{\odot}$) and 10 Myr ($0.012-0.08$ M$_{\odot}$) isochrones from
the DUSTY brown dwarf models of Chabrier et al. (2000).  These
model isochrones under-estimate the $I-$band magnitude by $1-2$
mag (and also under-estimate the colour) of low-mass stellar
members (M4-5 stars near $I = 14$) but approximately predict the
properties of objects near the brown dwarf -- planet boundary
(near $I = 18$). The solid-line is our assumed locus of $\sim 10$
Myr-old objects to select low-mass star and brown dwarf candidates
of the $\eta$ Cha cluster.  An apparent candidate at $I \approx
14$ and $(I-J) \approx 2.0$ is a known M4.7 member, ECHA
J0841.5--7853, with WFI/IRIS2 photometry that differs only
slightly from already-published values.  There are no other
obvious new candidate cluster members.

To see if we missed any other objects of interest, we checked all red
$(J-H)$ sources in the $(J-H)$ versus $J$ colour-magnitude diagram
(Fig. 4).  Objects plotted as crosses are those that do not appear
in the WFI $I$-band images, while those plotted as empty-plus signs
are saturated in the $J$-band images (with $J \gtrsim 13$).  As in
Fig. 3, the solid line represents our criterion for cluster candidates
in this colour-magnitude plane.  Inspection of images of the three
objects marked as crosses that reside close to the solid line with
$J \approx 16$ reveals them to be galaxies.

The completeness limit of the photometry which is the peak magnitude
of the luminosity function are $I = 19.1$, $J = 18.2$, and $H = 17.6$,
respectively.  These completeness limits are generally about 1 mag
brighter than the magnitude of the faintest stars measured.

\section{Results}

\begin{figure}
\begin{center}
\includegraphics[width=80mm]{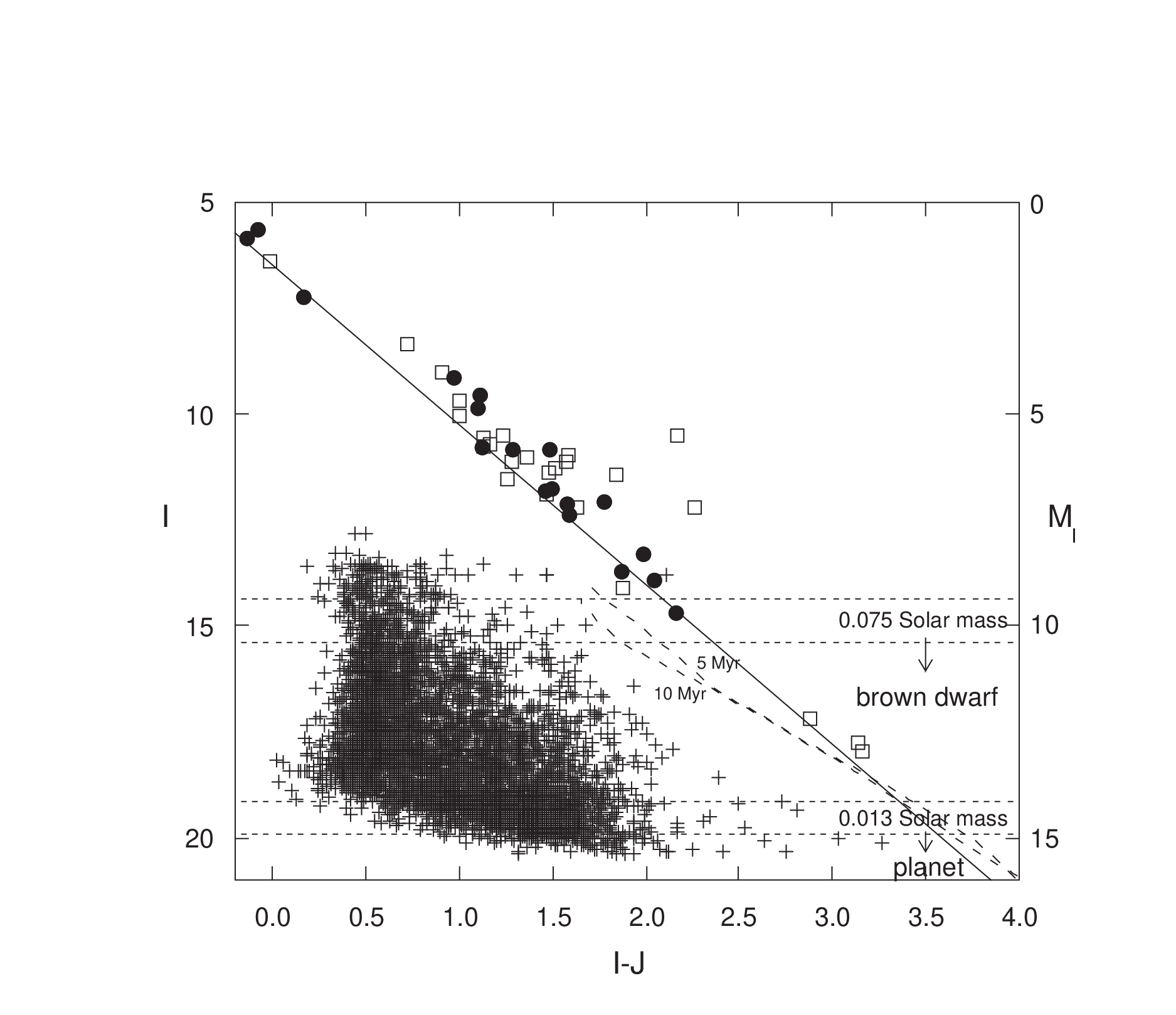}
%\vspace{-3cm}
\caption{The $I$ versus $(I-J)$ colour-magnitude
diagram for stars in the $\approx 33 \times 33$ arcmin surveyed
region of the $\eta$ Cha cluster. The right-hand-side axis is
marked in absolute magnitude. The filled-circles are for the known
members of the $\eta$ Cha cluster.  Open-squares are for the
members of the TWA if placed at the $\eta$ Cha distance (see
Mamajek 2005 for distances). The scatter in magnitude at any given
colour is principally a function of unaccounted for multiplicity.
The two long-dashed lines are the 5 Myr ($0.008-0.075$
M$_{\odot}$) and 10 Myr ($0.012-0.08$ M$_{\odot}$) DUSTY model
isochrones of Chabrier et al. (2000). The solid-line is our
criterion to choose low-mass star/brown dwarf candidates. The
horizontal short-dashed lines are the 0.075 M$_{\odot}$ and 0.013
M$_{\odot}$ brown dwarf mass boundaries derived from the mass and
$K$-band luminosity relations given within the 5 Myr and 10 Myr
DUSTY models.  We estimated the $I$-band magnitudes corresponding
to this brown dwarf mass range using the observed $IK$
colour-magnitude relations shown in Fig. 1.  These relations show
our survey is sensitive to brown dwarf masses down to the $\approx
0.013$ M$_{Jup}$ brown dwarf -- planet boundary.}
%\label{Fig3}
\end{center}
\end{figure}

\begin{figure}
\begin{center}
\includegraphics[width=80mm]{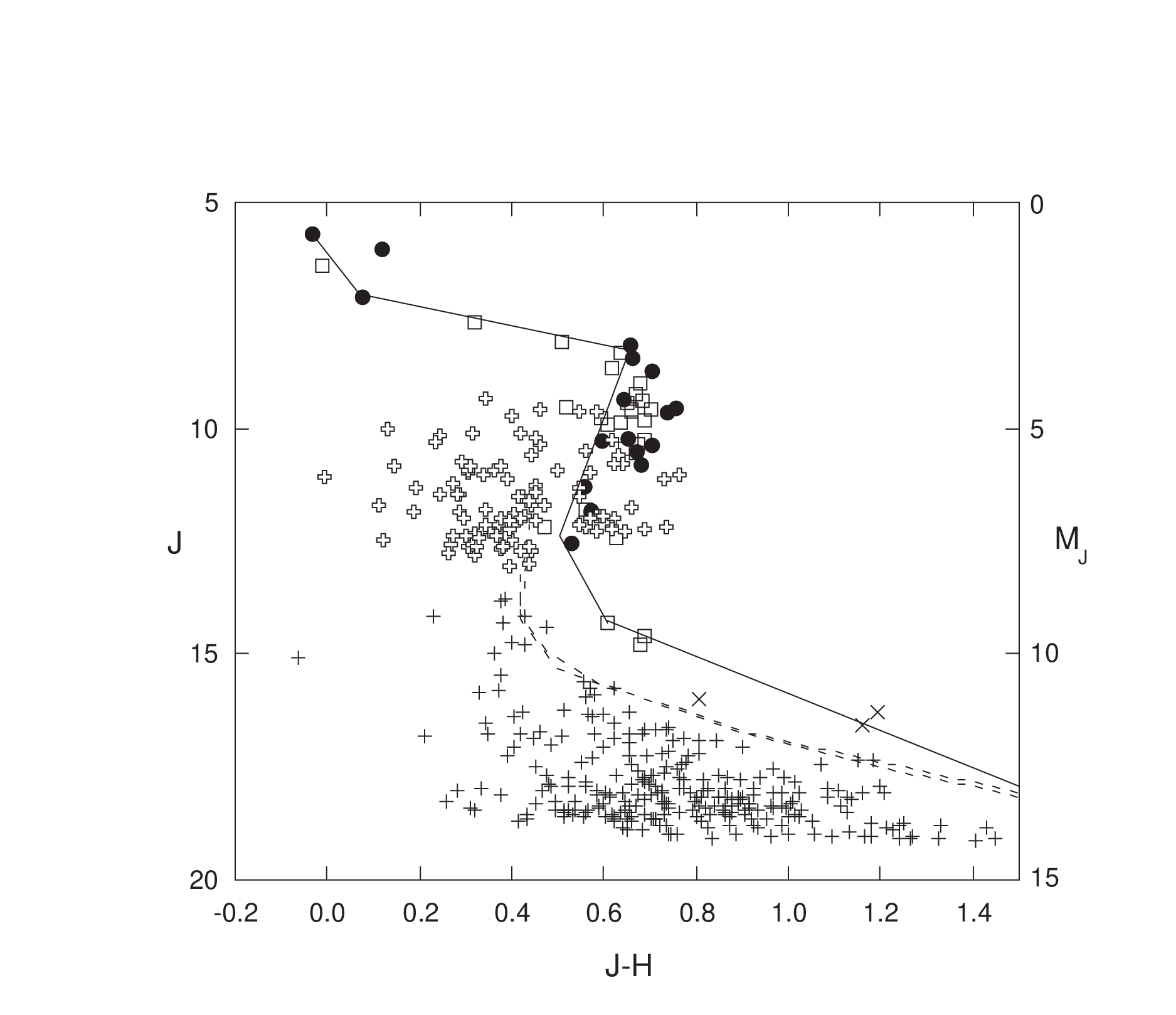}
%\vspace{-3cm}
\caption{The $J$ versus $(J-H)$ colour-magnitude
diagram for the central regions of $\eta$ Cha cluster from our
IRIS2 $JH$ data. The right-hand-side axis is marked in absolute
magnitude. Symbols have the same meaning as in Fig. 3 except that
crosses and empty-plus symbols denote undetected or saturated
objects, respectively, in the WFI $I$-band images.}
%\label{Fig4}
\end{center}
\end{figure}

We report the outcome of a deep optical and infrared photometry survey
of the central regions of the $\eta$ Cha star cluster ($33 \times 33$
arcmin), corresponding to an area of 0.9 pc$^{2}$ at the distance to
the cluster of $d = 97$ pc.  The completeness of the photometric data
implies that our survey could reveal members down to the brown dwarf
-- planet boundary of $\sim 0.013$ M$_{\odot}$, but we found no such
low mass cluster candidate members.

This null result indicates that either low mass objects were not
created in the $\eta$ Cha cluster or, if they were created, then
they have been dynamically scattered outward after their birth.
The former would suggest an unusual IMF for the cluster, highly
deficient in both low-mass stars and brown dwarfs.  This result
would be surprising given the apparent near-universality of the
IMF, unless sparse star clusters such as $\eta$ Cha are unusual as
our null result may suggest. Dynamical scattering of low mass
stars may still be plausible despite the fact that Luhman (2004)
did not find any low-mass candidate of the $\eta$ Cha cluster
within a distance of $1.5^{\circ}$ from the cluster center. The
present compact state of the cluster suggests that it is not
dynamically relaxed, and hints at its early dynamical history.
Adopting values from Lyo et al. (2004), the observed half-mass ($M
\approx 10$ M$_{\odot}$) radius of $r \approx 5$ arcmin = 0.15 pc
yields a {\it current\,} dynamical crossing time of $t_{ct}
\approx 600,000$ yr; a timescale that may have been considerably
shorter in the past if the cluster was more-massive (contained
more members) and/or physically more-compact. For an age of $t
\sim 10$ Myr, this calculation suggests the $\eta$ Cha cluster is
dynamically moderately-evolved, with the core having persisted for
a few tens of crossing times. Stars ejected from the cluster
during the first few $t_{ct}$ need only be imparted a velocity of
$\gtrsim 0.3$ km\,s$^{-1}$ to be removed $r > 1.5^{\circ}$ = 2.5
pc from the core. The possibility remains that a halo population
of low mass objects at radii of several degrees surrounds the
observed remnant cluster core.

\section*{Acknowledgments}

We thank the Anglo-Australian Observatory for the award of service
time, and Chris Tinny for our observation. WAL thanks UNSW@ADFA
FRG and SRG grant schemes for research support.

\bsp

\label{lastpage}

\end{document}